\date{}
\documentclass[showpacs]{revtex4}
\usepackage{graphicx,psfrag,amsmath,amssymb,amsfonts,bbm,latexsym,color,dcolumn,epsf}

\begin{document}
\title{Decoherence induced by a fluctuating Aharonov-Casher phase}

\author{Fernando C. Lombardo \footnote{lombardo@df.uba.ar}}
\author{Francisco D. Mazzitelli \footnote{fmazzi@df.uba.ar}}
\author{Paula I. Villar \footnote{paula@df.uba.ar}}
\affiliation{Departamento de F\'\i sica {\it Juan Jos\'e Giambiagi}, FCEyN UBA,
Facultad de Ciencias Exactas y Naturales, Ciudad Universitaria,
Pabell\' on I, 1428 Buenos Aires, Argentina}

\date{today}

\begin{abstract}
Dipoles interference is studied when atomic systems are coupled to
classical electromagnetic fields. The interaction between the
dipoles and the classical fields induces a time-varying
Aharonov-Casher phase. Averaging over the phase  generates a
suppression of fringe visibility in the interference pattern. We
show that, for suitable experimental conditions, the loss of contrast for dipoles
can be observable and
almost as large as the corresponding one for coherent electrons.
We analyze different trajectories in order to
show the dependence of the decoherence factor with the velocity of the
particles.
\end{abstract}


\maketitle

\newcommand{\beq}{\begin{equation}}
\newcommand{\eeq}{\end{equation}}
\newcommand{\dalam}{\nabla^2-\partial_t^2}
\newcommand{\mbf}{\mathbf}
\newcommand{\itm}{\mathit}
\newcommand{\beqa}{\begin{eqnarray}}
\newcommand{\eeqa}{\end{eqnarray}}


\section{Introduction}

Interference effects are the most distinguished characteristic of
quantum mechanics. The double-slit experiment is often used as the
starting point to quantum description of Nature. There, when the
path of the interfering particles is measured, the interference
pattern disappears and the classical picture is recovered. The
interaction between the quantum system with its environment
(measuring apparatus) is responsible for the process of
decoherence, which is one of the main ingredients in order to
reach the quantum-to-classical transition.

The Aharonov-Bohm (AB) \cite{aharonov} interference experiment
can be  a realistic
probe for the predictions of decoherence. This experiment starts
by the preparation of two electron wave packets $\varphi_1(\vec
x)$ and $\varphi_2(\vec x)$ in a coherent superposition, assuming
each of the charged particles follows a well defined classical path
($C_1$ and $C_2$, respectively). The complete wave function
involves the presence of the environment, given an state \beq \psi
(t = 0)=  \left[\varphi_1(\vec x) + \varphi_2(\vec x)\right]
\otimes \chi_0(\vec y), \eeq where $\chi_0(\vec y)$ represents the
initial quantum state of the environment (whose set of coordinates
is denoted by $\vec y$). As time follows, the electron's coherent
state entangles with the environment, and the total wave function
can be described as \beq \psi (t) =  \varphi_1(\vec x,t) \otimes
\chi_1 (\vec y,t) +
             \varphi_2(\vec x,t) \otimes \chi_2 (\vec y,t).
\eeq Therefore, the two delocalized electron states $\varphi_1$
and $\varphi_2$ become correlated with two different states of
the environment. The probability of finding a particle at a given
position at time $t$ (for example when interference pattern is
examined) is, \beq \mbox{Prob}(\vec x,t) = \vert \varphi_1(\vec
x,t)\vert^2 + \vert \varphi_2(\vec x,t)\vert^2 + 2 {\rm Re}\left(
\varphi_1(\vec x,t) \varphi^*_2(\vec x,t) \int d^3y ~\chi_1^*(\vec
y,t)\chi_2(\vec y,t)\right). \eeq

The overlap factor $F = \int d^3y ~\chi_1^*(\vec y,t)\chi_2(\vec
y,t)$ is responsible for two separate effects. Its phase
generates a shift of the interference fringes, and its absolute
value is responsible for the decay in the interference fringe
contrast. Of course, in absence of environment the overlap factor
is not present in the interference term. When the two
environmental states do not overlap at all, the final state of
the bath identifies the path the electron followed. There is no
uncertainty respect to the path. Decoherence appears as soon as
the two interfering partial waves shift the environment into
states orthogonal to each other. Last statement suggests that the
environment saves (in some way) the information about the path
the electron takes.

The loss of quantum coherence can alternatively be explained by
the effect of the environment over the partial waves, rather than
how the waves affect the environment. As has been noted, when a
static potential $V(x)$ is exerted on one of the partial waves,
this wave acquires a phase, \beq \phi = - \int V[x(t)] dt, \eeq
and therefore, the interference term appears multiplied by a
factor $e^{i\phi}$. This is a possible agent of decoherence. The
effect can be directly related to the statistical character of
$\phi$, in particular in situations where the potential is not
static. Even more, any source of stochastic noise would create a
decaying coefficient. For a general case, $\phi$ is not totally
defined, i.e. it is described by means of a distribution function
$P(\phi )$. From this statistical point of view, the phase can be
written as \beq \langle e^{i\phi}\rangle = \int e^{i\phi} P(\phi
) d\phi.\label{IF} \eeq

In this way, the uncertainty in the phase produces a decaying
term that tends to eliminate the interference pattern. This
dephasing is due to the presence of a noisy environment coupled
to the system and can be also represented by the Feynmann-Vernon
influence functional formalism. It is easy to prove that
Eq.(\ref{IF}) is the influence functional generated after
integrating out the environmental degrees of freedom of an open
quantum system. Therefore, in Ref. \cite{stern} was shown the
formal equivalence between the two ways of studying dephasing
\beq \langle e^{i\phi}\rangle = F =  \int d^3y  ~ \chi_1^*(\vec
y,t)\chi_2(\vec y,t). \label{overlap}\eeq The overlap factor $F$
encodes the information about the statistical nature of noise.
Therefore, noise (classical or quantum) makes $F$ less than one,
and the idea is to quantify how it
slightly destroys particle interference pattern.

In many cases, the interaction with the environment cannot be
switched off. Thus, for charged particles or neutral atoms with
dipole moment, the interaction with the electromagnetic field is
crucial. This interaction induces a reduction of fringe
visibility. For example, vacuum fluctuations of electromagnetic
fields have been considered as a decoherent agent \cite{villa}.
Double-slit-like experiments were studied in the presence of
conductors, which change the structure of vacuum and modify
predictions about decoherence effects. The changes on the fringe
visibility induced by the position and relative orientation of the
conductors are rather small.

Electron interference in mesoscopic devices irradiated by external
nonclassical microwaves was considered in Ref. \cite{vourdasPRA02}.
The effect of quantum noise on electron interference for several
types of microwaves has been analyzed, and it was shown that
entangled electromagnetic fields interacting with electrons
produce entangled photons \cite{vourdasPRA04}. Despite the quantum
nature of noise (vacuum fluctuations for example), classical noise
is also present when considering time dependent fields, or some
random variables that parametrize the environment. The destruction
of electron interference by external classical and quantum noise
has been studied in Ref. \cite{vourdasPRA01}.

In Ref. \cite{ford}, the overlap factor $F$ was evaluated
from a different point
of view. They studied the effect of time-varying electromagnetic
fields on electron coherence, including the statistical origin of
the AB phase $\phi$. However, they didn't consider $\phi$ neither
as coming from quantum fluctuations nor a time dependent field.
They included a random variable $t_0$, which was defined as the
electron emission time. This variable produces a fluctuating phase
$\phi$, and an average over it is needed in order to obtain the
result of the double-slit interference experiment. In this simple
version of decoherence, the role of a quantum environment is
replaced by a time-dependent external field which gives a
time-varying AB phase. The authors have considered the case of a
linearly polarized, monochromatic electromagnetic wave,
propagating in a direction orthogonal to the plane containing two
electron beams. The effect on the fringes seems to be
sufficiently large to be observable.

In the present article we follow last idea. We evaluate the
overlap factor $F$ for coherent neutral particles with permanent
(electric and magnetic) dipoles that are affected by time-varying
external fields. We consider two different cases with possible
experimental interest: firstly, the case of dipoles interacting
with a linearly polarized electromagnetic plane wave. This will
generalize results of Ref. \cite{ford} to the case of
Aharonov-Casher (AC) \cite{casher} case in atomic systems
\cite{sangster}, where coherent dipoles follow a closed path
around an external field. Secondly, the case of dipoles travelling through a
waveguide with rectangular section. As we will discuss, not all
these effects foresee an
observable displacement on the interference pattern related to
the phase shift of the wave function of the system. Dependence
on the velocity of the interfering particle is crucial.
We will show different configurations where the decoherence factor
depends on the particle's speed in a different way.  

In principle, for neutral particles with the same mass and velocity than
the charged ones, one should expect the loss of contrast for dipoles
to be smaller than the analogous one in experiments performed
using charged particles. Although this is basically true, the
interaction between electromagnetic fields and dipoles is also
weaker, allowing to increase the intensity of external classical
fields in the same amount than the dipole effect is smaller,
still neglecting the dipole scattering in the external field.
Therefore, for sufficiently strong external fields, the effect on
fringe visibility for dipoles will be of comparable magnitude
with respect to the case of charged particles. In order to prove this,
we will evaluate the scattering cross section for dipoles in
interaction with both a plane wave and the fields inside a
waveguide.

The paper is organized as follows. In Section II we present the
evaluation of the phase shift and the decoherence factor for
coherent dipoles (electric and magnetic) in the presence of a time
varying electromagnetic plane wave. Symmetric and non-symmetric trajectories
are considered in order to analize the dependence on the
velocity of the interfering particles. We also include in this section
some new results for charged particles. Section III shows the results
when the coherent trajectories are inside a waveguide. In Section
IV we estimate the decoherence factor for realistic values of the different
parameters, and
compare the loss of contrast for charged particles and dipoles.
Section V contains our final remarks. We also
include an Appendix where we compute the Thomson cross section
for dipoles.


\section{Coherent dipoles and a plane wave}

The AB phase \cite{aharonov}, known to arise when two coherent
electrons traverse two different paths $C_1$ and $C_2$
in the presence of an
electromagnetic field, is ($c=\hbar=1$) \beq \phi=-e \oint_{
\delta \Omega}  dx_{\nu} A^{\nu}(x), \label{faseAB} \eeq where
$\delta \Omega =C_1-C_2$ is a closed spacetime path. If the electromagnetic
field's fluctuations happen in a time scale shorter than the total
time of the experiment, this shift of phase results in a loss of
contrast in the interference fringes. Then, the overlap factor (or
decoherence factor) is given by Eq.(\ref{overlap})
\cite{stern,ford}. In that expression, angular brackets denote
either an ensemble of quantum noise or a time average over a
random variable.

The classical and quantum interaction of a dipole with an
arbitrary electromagnetic field has been described in detail in
Ref. \cite{anandan}, where a unified and fully relativistic
treatment of this interaction has been presented. The
Lorentz-invariant, classical interaction Lagrangian is
$\frac{1}{2}P_{\mu\nu}F^{\mu\nu}$ where $P_{\mu\nu}$ is the
antisymmetric dipole tensor \cite{villa, anandan}. In the rest
frame of the particle, the electric ($\mbf{d}$) and magnetic
($\mbf{m}$) dipole moments can be obtained from $P_{0i}= d_i$ and
$P_{ij}=\epsilon_{ijk}m_k$ respectively.

In the quantum case, the phase shift that two neutral particles
with electric and magnetic dipole moments experience due to a
classical time dependent electromagnetic field is known as the
Aharonov-Casher phase \cite{casher} and is defined by \beq \phi=-
\oint_{\delta \Omega} a_{\nu}(x) dx^{\nu}, \label{faseAC} \eeq
where $a_{\nu}(x)=(-\mbf{m} \cdot \mbf{B} -\mbf{d} \cdot \mbf{E},
\mbf{d} \times \mbf{B} - \mbf{m} \times \mbf{E})$ plays the role
of $A_\nu$ in the AB case, $\delta \Omega=C_1-C_2$, and
$C_1$ and $C_2$ are the two paths followed by
the particles that interfere.

In order to evaluate the integral in Eq.(\ref{faseAC}), we
consider the case of a linearly polarized monochromatic wave of
frequency $\omega$ propagating in the $\hat{y}$ direction, with
an electric and magnetic field in the $\hat{z}$ and $\hat{x}$
direction respectively. We will also assume that the particles' path
is
confined to the $\itm{{\hat x}-{\hat z}}$ plane (see Fig. 1).
We can
write the plane wave as $\mbf{E}(x)=E_0 \sin(wt-ky)~\hat{z}$,
$\mbf{B}(x)=E_0 \sin(wt-ky)~\hat{x}$ and compute $a_\nu$, which
is given by

\beqa
a_{\nu}(x)&=&(-d_z E_z-
m_x B_x, m_y E_z,d_z B_x-m_x E_z,-d_y B_x)\nonumber \\
&=& E_0 (-d_z - m_x , m_y ,d_z - m_x,-d_y) \sin (\omega t - k y)
\nonumber\\ &\equiv &{\tilde a}_{\nu}\sin (\omega t - k y).
\label{anu}
\eeqa

Following Ref. \cite{ford}, we will assume that the phase $\phi$
depends on a random variable $\xi = \omega t_0$ given by the
emission time of the dipoles. It is the time $t_0$ at
which the center of a localized wave packet is emitted. When the
measuring time takes longer than the flight time, we will observe
a result which is the temporal average over $t_0$. Thus, $t_0$ is
a random variable by which $\phi$ has to be averaged. We can
write the AC phase $\phi$ as
\beq \phi (t_0) =-\oint_{\delta\Omega} {\tilde a}_\nu \sin
(\omega t - k y + \omega t_0) ~ dx^\nu, \eeq before taking the
time average, we rewrite the phase as \beq \phi (t_0) = A \cos
(\omega t_0) + B \sin (\omega t_0), \eeq where \beqa A &=&
-\oint_{\delta\Omega} {\tilde a}_\nu(x) \sin (\omega t - k
y)~dx^\nu,
\nonumber \\
B &=& -\oint_{\delta\Omega} {\tilde a}_\nu \cos (\omega t - k y)~dx^\nu .
\label{AyB}\eeqa
The average over the random phase (generating a classical noise) produces a
decoherence factor
\beq
F = \langle e^{i\phi}\rangle = \lim_{T\rightarrow \infty}\frac{1}{2T}
\int_{-T}^{T}dt_0 \exp\left\{i\left[A \cos (\omega t_0) + B \sin (\omega t_0)
\right]\right\}=
J_0(\vert C\vert ),
\eeq
where $J_0$ is the Bessel function. The modulus of
complex number $C = A + i B$ measures
degree of decoherence. The overlap factor $F$ decreases from
one to zero as $\vert C\vert$ varies between zero and the
first zero of $J_0$. For larger values of $\vert C\vert$, the
overlap factor oscillates with decreasing amplitude.
For a Gaussian $P(\phi)$ distribution with
$\langle \phi^2 \rangle \ll 1$, we have, in the limit
$\vert C\vert \ll 1$, $\langle e^{i\phi}\rangle \approx 1
- \langle \phi^2\rangle = 1 - \vert C\vert^2/2$.

A characteristic feature of the usual  AB and AC effects is that
the phase shift is independent of the velocity \cite{berry} of the
particle and there is no force on the particle \cite{APV}.
Moreover, the phase shift depends only on the topology of the
closed spacetime path $\delta \Omega$. Of course, these properties
are not longer valid when  the external field is time dependent because
the particle does suffer a net force applied on it.
 Thus, in order to analyze the dependence upon the trajectory, we will
evaluate Eq.(\ref{faseAC}) for different paths and will find that the phase's
dependence on the velocity is strongly related to the trajectory the particles
follow.

\subsection{Symmetric trajectories}

\begin{figure}[!h]
\includegraphics[width=10cm]{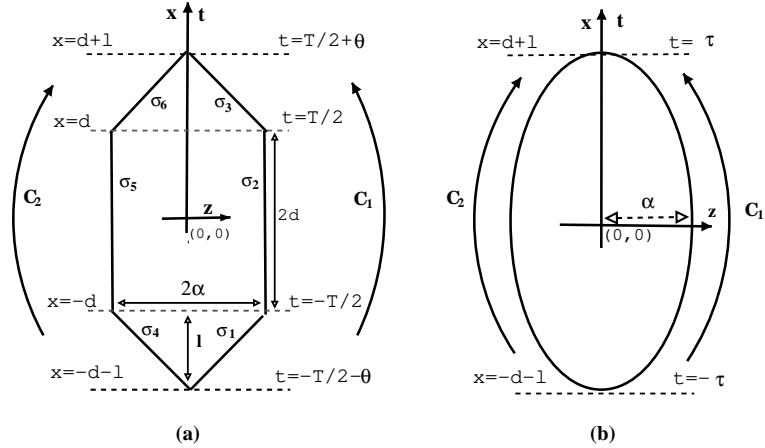}
\caption{Paths ${\cal C}_1$ and ${\cal C}_2$ are shown for the (a) trajectory used
in \cite{ford} and (b) an elliptic one.}
\label{fig1}
\end{figure}

In this subsection, we will estimate the phase
acquired by two neutral particles with electric and magnetic
dipole moments when they follow symmetric
trajectories as the ones depicted
in Fig. \ref{fig1}.

To begin with, we will consider the same trajectory used
by authors in \cite{ford} to analyze the AB
phase factor between charged particles.
To perform integration in Eq.(\ref{AyB}) along the
trajectory of Fig.1(a), we must calculate

\beq \phi=\oint_{\delta \Omega} a_{\nu}(x) dx^{\nu}= \sum_{i=1}^6
\int_0^1 a_{\nu} (\sigma_i^{\nu}(u))\cdot \frac{d\sigma_i ^{\nu}}{d u}
\label{parametrize}
du, \eeq where $\sigma_i$,  with $i=1...6$, parametrize
the different
segments of the trajectory.
The path of particle 1 (${\cal C}_1$) is
described by:
\begin{eqnarray}
\sigma_{1}(u) & = & (-T/2-\theta + \theta u,-d-l+u l,0,u \alpha)  \nonumber \\
\sigma_{2}(u) & = & (-T/2 + Tu,-d+2du,0,\alpha) \nonumber \\
\sigma_{3}(u) & = & (T/2+u\theta, d+ul,0,\alpha -u \alpha),
\quad \quad \mathrm{for} \quad
0 \leq u\leq 1.
\label{param-sym}
\end{eqnarray}
Meanwhile the path of particle 2 (${\cal C}_2$) is:
\begin{eqnarray}
\sigma_{4}(u) & = & (-T/2-\theta + \theta u,-d-l+u l,0,-u \alpha) \nonumber \\
\sigma_{5}(u) & = & (-T/2+T u,-d+2 d u,0,- \alpha) \nonumber \\
\sigma_{6}(u) & = & (T/2+u \theta, d+ul,0,-\alpha +u \alpha), \quad \mathrm{for} \quad
0 \leq u\leq 1.
\end{eqnarray}

Performing the integrations in Eq.(\ref{parametrize})
and using the definition given in Eq.(\ref{AyB}), we obtain $B_d=0$ and
\beq
\vert C_d\vert\equiv
\vert A_d \vert = 4 E_0 d_y \big(\frac{ 2 \alpha}{\omega \theta} \big)
\sin \big(\frac{\omega \theta}{2}\big) \sin \big(\frac{\omega (T +
\theta)}{2} \big),
\label{Cd}
\eeq where $2 \alpha$ is the
maximum distance between the particles, $d_y$ is the electric
dipole's moment in the $\hat y$ direction and $T,\theta$ are
characteristic times of the trajectory. The subindex $d$ indicates
that these are the values of $A,B$ and $C$ for dipoles.

For non relativistic particles, we expect
$\omega\theta,\,\,\omega T\gg 1$. Therefore,
in order to have an estimation for $C_d$ we can replace
the sine functions by a typical value $1/\sqrt 2$.
In order to make explicit the dependence on the
velocity, note that $\theta= s/ v$, where $s$ is the
length of the first and third segments of the path defined by
$s =\sqrt{\alpha^2+ l^2}$. Using this, we rewrite
Eq.(\ref{Cd}) as
\beq \vert C_d\vert \approx
\frac{2}{\pi}E_0
d_y \big( \frac{\alpha}{s} \big)\lambda  v \approx
\frac{2}{\pi} e E_0
\big( \frac{\alpha}{s} \big)\lambda L  v
. \label{Cdbis} \eeq
Here $L$ is the characteristic length
of an atom with electric dipole $d= e L$ ($L \approx 10^{-9} m$),
and $\lambda$
is the wavelength of the plane wave.
The analogous result for charged particles is given by
 $|C_e|\approx \frac{1}{\pi^2} e E_0 \lambda^2 (\frac{
\alpha}{s}) v$ \cite{ford}. Assuming that the
charged and neutral particles have the
same speed and trajectory, for
the same external field we obtain
\beq |C_d|
\approx |C_e| \bigg(\frac{L}{\lambda}\bigg). \label{comparacion} \eeq

Even though Eq.(\ref{comparacion}) might be discouraging, the scattering
cross section for neutral particles is much
smaller than for charged
particles (see Appendix). This makes it possible to increase the
intensity
of the external field and, consequently, the value of the
decoherence factor at least some orders of magnitude.

If the neutral particles follow
an elliptic path
(Fig \ref{fig1} (b)), the calculations proceed in a similar way.
The trajectory is parametrized by
\beqa
\sigma_1(u) = (\tau \sin(u),(d+l) \sin(u),0,~~\alpha \cos(u))
 \quad \quad \quad \mathrm{for} -\pi/2 \leq u \leq \pi/2 \nonumber \\
\sigma_2(u) = (\tau \sin(u),(d+l) \sin(u),0,-\alpha \cos(u))
 \quad \quad \quad \mathrm{for} -\pi/2 \leq u \leq \pi/2,
\label{param-ellip}
\eeqa
where $\tau$ is the time of flight of the dipoles and $(d +l)$
is the total length of the path. In this case, the quantity
$|C^d_{\rm ellip}|$ is
given by
\beq |C^d_{\rm ellip}|= 2 \pi \alpha E_0 d_y
\mathrm{J}_1 [\omega \tau], \eeq with $\mathrm{J}_1$ the Bessel
function of first order.
Using the asymptotic expansion of this function for $\omega \tau \gg 1$
we find
\begin{equation}
|C^d_{\rm ellip}| \approx \frac{\sqrt 2\pi\alpha E_0d_y}{(\omega \tau )^{1/2}}
= \sqrt \pi\alpha e E_0 L \left(\frac{v \lambda}{s'}\right)^{1/2}
\label{Cd_ellip} \end{equation} where $s'$ is the length travelled
by the neutral particles at a speed $v$ and in a time $\tau$.
It is
important to note that while $\vert C_d\vert$ in Eq.(\ref{Cdbis})
depends linearly on the velocity, for the elliptic
trajectory $|C^d_{\rm ellip}|$ scales as $\sqrt v$.

It is
interesting to check whether the same
difference in behaviours applies to the case of the AB phase for
charged particles or not. If the charged particles travel across
the paths shown in Fig.1(b),
the quantity $|C^e_{\rm ellip}|$ can be computed
from Eq.(\ref{faseAB}) and the parametrization Eq.(\ref{param-ellip}).
The result is
\beq |C^e_{\rm
ellip}|=  2 \pi \alpha e E_0 \lambda \mathrm{J}_1[\omega \tau], \eeq
showing that the dependence on the velocity is
similar for both neutral and charged particles.

Finally, it is worth noting that, as both trajectories  in
Figs.1(a) and (b) are symmetric respect to the $\hat x$ ($\hat t$) axis,
only the term proportional to $A$ of Eq.(\ref{AyB}) contributes to the AC
phase, and,
consequently, to the decoherence factor. This is due to the parity of
the integrand with respect to $x$ and $t$.

\subsection{Asymmetric trajectory}

In this subsection we will consider the case of dipole
wave packets travelling
across the asymmetric trajectory depicted
in Fig. 2. As we will see, in this case, not
only do both terms
in Eq.(\ref{AyB}) contribute to the AC phase but also
different components of the dipole moments. Consequently, speed dependence will
be different from the case of symmetric trajectories.

\begin{figure}[!h]
\includegraphics[width=5cm]{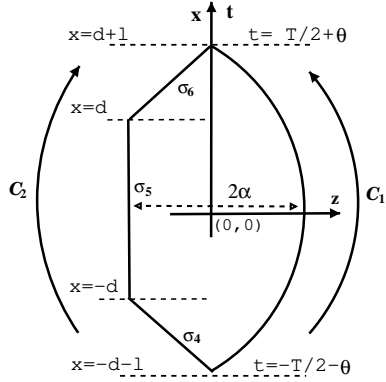}
\caption{Paths ${\cal C}_1$ and ${\cal C}_2$ are shown for the asymmetric trajectory.}
 \label{fig2}
\end{figure}

We write $C^d_{\rm asym}= A^d_{\rm asym} + i B^d_{\rm asym}$,
where each of these coefficients is defined in Eq.(\ref{AyB}).
The parametrization of the asymmetric curve can be read from
Eqs.(\ref{param-sym}) and (\ref{param-ellip}).
After
performing the corresponding integrations, we obtain \beqa A^d_{\rm
asym} &=&\pi d_y E_0 \alpha \mathrm{J}_1[\omega \tau] +
\frac{4E_0d_y \alpha}{\omega \theta} \sin [\frac{\omega\theta}{2}]
\sin [\frac{\omega}{2}(T + \theta)],\nonumber \\
B^d_{\rm asym} &= & (d_z + m_x)\frac{2E_0}{\omega} \left[
\sin [\frac{\omega \theta}{2}] \cos[\frac{\omega}{2}(T + \theta)] +
 \sin[\frac{\omega T}{2}] \right] \nonumber \\
&+&  \frac{2 m_y E_0 l}{\omega \theta} \sin
[\frac{\omega \theta}{2}] \cos[\frac{\omega}{2}(T+\theta)] +
\frac{4 d E_0 m_y}{\omega T} \sin[\frac{\omega T}{2}]
.\eeqa

The first term in the $B$ coefficient is the most important
in the low velocity
limit, because all other terms are ${\cal
O}(1/\sqrt{\omega \tau})$ or ${\cal O}(1/\omega \theta)$, which
are much less than unity. The quantity $|C^d_{\rm asym}|$ is then
dominated by that contribution and can be approximated
by
\beq |C^d_{\rm asym}| \approx \frac{e}{\pi}  E_0  L
\lambda\,\, , \label{Cd_asym} \eeq
being independent of the velocity.

For charged particles, the situation is very different.
The decoherence factor can be
computed by integrating Eq.(\ref{faseAB}) along the asymmetric
trajectory of Fig. \ref{fig2}. We obtain $C^e_{\rm asym}= A^e_{\rm asym} +
i B^e_{\rm asym}$ with
\beqa A^e_{\rm asym} &=& 2\pi e E_0 \frac{\alpha}{\omega}
\mathrm{J}_1[\omega \tau] + \frac{4 e E_0 \alpha}{\omega^2 \theta}
\sin [\frac{\omega\theta}{2}]
\sin [\frac{\omega}{2}(T + \theta)],\nonumber \\
B^e_{\rm asym} &= & 0
.\eeqa
Therefore, we can approximate
\beq
|C^e_{\rm asym}| \approx \frac{e}{ \sqrt{2\pi}}  E_0 \alpha 
\bigg(\frac{v\lambda^3}{s'}\bigg)^{1/2},\label{Ce_asym} \eeq
which depends on the velocity as for the case of the elliptic path.
What is worthy of note is the fact
that while this result does depend
on the velocity of the electrons, the decoherence factor for the
dipoles does not.


\section{Coherent dipoles inside a waveguide}

Now we consider the field generated inside a waveguide (along the
$\hat{y}$ direction) with rectangular section. For the TE mode,
the electromagnetic
fields inside the pipe are the real part of
\begin{eqnarray}
B_y & = & B_0 \cos(k_x x) \cos(k_z z) \exp(i(k_y y - \omega t)), \nonumber \\
B_x & = & \frac{-i k_y k_x}{\gamma^2} B_0 \sin(k_x x) \cos( k_z z)
 \exp(i(k_y y - \omega t)), \nonumber \\
B_z & = & \frac{-i k_y k_z}{\gamma^2} B_0 \cos(k_x x) \sin(k_z z)
\exp(i(k_y y - \omega t)), \nonumber \\
E_x & = & \frac{ i \omega k_z}{\gamma ^2} B_0 \cos(k_x x) \sin(k_z z)
\exp(i(k_y y - \omega t)), \nonumber \\
E_z & = & \frac{ i \omega k_x}{\gamma ^2} B_0 \sin(k_x x) \cos(k_z z)
\exp(i(k_y y - \omega t)),
\end{eqnarray}
where $k_x = \frac{ m \pi }{b}$, $k_z = \frac{ l \pi}{a}$ (with
$m$ and $l$ integers and $a,b$ the dimensions of the pipe),
$\gamma= \sqrt{ (\frac{ l \pi}{a})^2 + (\frac{m \pi}{b})^2}$,
$\omega \equiv k = \sqrt{\gamma^2+k_y^2}$, and $B_0$ is a complex
number. We will write $B_0$ as $B_0= |B_0| \exp (i \omega t_0)$,
where $t_0$ is the particle emission time, as was defined in the
previous Section.

\begin{figure}[!h]
\includegraphics[width=3cm]{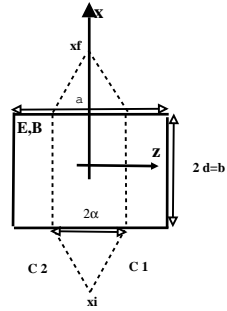}
\caption{ The particles interfere in the presence of a time-dependent electromagnetic
 field. The particles follow the path $C_1$ and $C_2$ in the $\itm{x-z}$
 plane as shown above for the waveguide.}
\label{guia}
\end{figure}

Let us consider the trajectory depicted in  Fig. 3. The particles
are set together at one side of the
pipe ($x_{\rm i}$) (at some initial time $t_0$), released and meet
again at the other side ($x_{\rm f}$),
having gone through it along straight paths. As the fields
vanish outside the waveguide, the particles will interact
with the electromagnetic field only when travelling inside
it. Bearing in mind that $C^d_{\rm guide} = A^d_{\rm guide}
+ i B^d_{\rm guide}$, we obtain for this case
\beqa
A^d_{\rm guide} & = & 0 \nonumber \\
B^d_{\rm guide} & = & -\frac{ 4
}{( \omega T)^2-( m \pi)^2} \vert B_0\vert \frac{k_z}{\gamma^2}
\sin (\frac{l \pi \alpha}{a}) \bigg[ \omega T \cos(\frac{ m
\pi}{2}) \sin(\frac{\omega T}{2})
\nonumber \\
& - & m \pi \cos(\frac{\omega T}{2}) \sin(\frac{ m \pi}{2}) \bigg]
\bigg( d_x \omega T - m_z k_z T- 2 d_y k_y \bigg),
\label{te general}
 \eeqa where $2d=b$ is the total distance travelled by the
dipoles inside the pipe, $b$ and $a$ are the dimensions of the
waveguide, $2 \alpha$ is again the maximum distance the dipoles are
moved apart, $m,l$ are the modes of the waveguide and $T$ the time
the dipoles are travelling inside the pipe.

For the first mode of the waveguide ($l=1,m=0$), this expression
looks simpler: \beq C^d_{\rm {TE_{10}}}= -\frac{4 }{\omega T} \vert
B_0\vert \frac{a}{\pi} \sin(\frac{\pi \alpha}{a})
\sin(\frac{\omega T}{2})
 (d_x \omega T-m_z k_y T-b d_y k_y).
\eeq
We can estimate the quantity $|C^d_{\rm {TE_{1,0}}}|$
in the same way we did in the preceding section, yielding
 \beq |C^d_{\rm {TE_{10}}}| \approx
\frac{2\sqrt 2}{\pi} B_0~a~\sin(\frac{\pi \alpha}{a})
d_x,\label{Cd_guia10} \eeq where we have assumed that
$|\mbf{d}| >> |\mbf{m}|$. This result is independent of
the dipoles'
velocity once more. Performing the same calculation with charged
particles,
 the corresponding $\vert C^e_{\rm {TE_{10}}}\vert$ factor reads
\beq
|C^e_{\rm {TE_{10}}}|\approx \frac{1}{(2 \pi^3)^{1/2}} e B_0 a \lambda
v \sin \big(\frac{\alpha \pi}{a} \big)
.\eeq
This result, opposite to the dipoles' one, depends on the
velocity of the charged particles.

Considering the dipoles again, from Eq.(\ref{te general})
we can compute the decoherence
factor for an arbitrary $\rm {TE}$ mode of the electromagnetic
field inside the cavity, $m \ne 0$ and $l \ne 0$, assuming $2 d=b$.
The result  is
\beqa C^d_{\rm TE} \approx - \frac{4 B_0}{(\omega
T)^2-(m \pi)^2} \frac{k_z}{\gamma^2} \sin(\frac{l \pi \alpha}{a})
\bigg[ \omega T \cos(\frac{m \pi}{2}) \sin(\frac{ \omega T}{2}) -
m \pi \cos(\frac{\omega T}{2}) \sin(\frac{m \pi}{2})\bigg]
\nonumber
\\\bigg(d_x \omega T -m_z k_z T- b d_y k_y \bigg) \sin(\omega t_0).
 \eeqa

It is easy to prove that the result is negligible in the case of
odd $m$ and $m\pi \ll \omega T$. In any other case we get
\beq
|C_{\rm TE}^d| \approx B_0 a \vert\sin (\frac{\pi\alpha}{a})\vert \bigg[
d_x^2 + (\frac{b}{\omega T}d_y k_z)^2 - \frac{2 b}{\omega T} d_x d_y
k_y\bigg]^{1/2}, \label{TEgeneral} \eeq
which has the same magnitude than for the lowest
$\rm {TE}_{10}$ mode  Eq.(\ref{Cd_guia10}).

For the TM modes, the fields inside the pipe are
\begin{eqnarray}
B_x & = & \frac{i\omega k_z}{\gamma^2} E_0 \cos(k_x x) \cos(k_z z)
\exp(i(k_y y - \omega t)), \nonumber \\
B_z & = & \frac{-i \omega k_x}{\gamma^2} E_0 \sin(k_z z) \cos( k_x x)
 \exp(i(k_y y - \omega t)), \nonumber \\
E_x & = & \frac{i k_y k_x}{\gamma^2} E_0 \cos(k_x x) \sin(k_z z)
\exp(i(k_y y - \omega t)), \nonumber \\
E_y & = & E_0 \sin(k_x x) \sin(k_z z)
\exp(i(k_y y - \omega t)), \nonumber \\
E_z & = & \frac{i k_y k_z}{\gamma^2} E_0 \sin(k_x x) \cos(k_z z)
\exp(i(k_y y - \omega t)).
\end{eqnarray}
where $E_0= |E_0| \exp (i \omega t_0)$. After some algebra, it is
possible to show that \beqa C^d_{\rm TM}= \frac{ -4}{( \omega T)^2-( m
\pi)^2} \vert E_0\vert \sin (\frac{l \pi \alpha}{a}) \bigg[ k_x
\bigg( \omega T \cos(\frac{ m \pi}{2})
\sin(\frac{\omega T}{2}) \nonumber \\
- m \pi \cos(\frac{\omega T}{2}) \sin(\frac{ m \pi}{2}) \bigg)
\bigg( -d_x \frac{k_y}{\gamma^2} T + m_z \frac{\omega}{\gamma^2}T+
b d_y \frac{\omega}{\gamma^2} \bigg)+ \nonumber \\
\bigg( \omega T \cos(\frac{\omega T}{2}) \sin(\frac{m \pi}{2}) - m
\pi \cos(\frac{m \pi}{2}) \sin(\frac{\omega T}{2}) \bigg) \bigg(-d_y
T-bm_z \bigg) \bigg] \sin(\omega t_0). \eeqa
Using the same arguments than in the TE case, we can check that
there is also a contribution independent of the dipoles' velocity
but this time proportional to $m_z$. Therefore, as we are
assuming $|\mathbf{m}|<<
|\mathbf{d}|$ all along this work, we conclude that $|C_{\rm TM}^d| \ll
|C_{\rm TE}^d|$.

It is worth noting  that in the TM case, as there is no component of the
magnetic
field along the guide axis, the AB phase vanishes. Therefore, the effect
for neutral particles is
the only one we expect.

\section{Numerical estimations}

When we compared the decoherence factors for charged and neutral
particles (see for example Eq.(\ref{comparacion})), we had assumed
that the relevant parameters (velocity, maximum separation) were the
same in both cases.
Even
though it is a valid thread of thought, it is not a realistic one.
Therefore, in this section we will estimate the loss of
coherence in more realistic situations.

In interference experiments with electrons, the wave packets can
be moved apart up to $100 \mu\rm{m}$ \cite{hasselbach}.
A typical non relativistic velocity  is $v_e \sim 0.1$. This
yields a relation $\omega T \sim 10$ for a field that has a
wavelength of about $100 \mu\rm{m}$. On the other hand,
in atomic interferometry, two neutral
particles can be separated up to $1~\rm{mm}$
\cite{keith,pfau}. Typical speeds are of the order
$v_d \sim 10^{-5}$
\cite{green}. We will assume the dimensions of the pipe $a\sim\rm{cm}$.

The energy density of the electromagnetic
field is defined as $\rho_{\rm
wave}= E_0^2/2$ for a plane wave. The energy density for the field
inside a waveguide is $\rho_{\rm guide}= (a B_0)^2/\lambda^2$.
Considering the typical values of $\lambda$ and $a$ given above,
we note that $a/\lambda \sim 1$ for dipoles whereas $a/\lambda \sim
10$ for electrons. As we
we are using Lorentz Heaviside units with $\hbar=c=1$, $\rho$ is also
the energy flux in the electromagnetic wave. We will
assume an energy flux of 10 $\rm{Watts}/\rm{cm}^2$,
approximately.

With all these values, we can estimate the
$C$ factor for all the cases presented in the previous sections.
The results are summarized in Table 1. As we can see, all the
results for electrons are of order one  or bigger,
which means that the effect is
experimentally observable.
Dipoles' results are smaller but not
that much as one would naively expect.

\begin{table}[b]
\begin{tabular}{|c|c|c|}
  \hline
& &\\
\textbf{Trajectories} & \textbf{Electrons}& \textbf{Dipoles}\\
  \hline \hline
  & & \\
   $~~|C|~~~ $ & $1$  &  $ 10^{-6} $\\ \hline
   & &\\
  $|C_{\rm ellip}|~$& $ 10$  & $
  10^{-3}$ \\
  \hline
   & & \\
   $|C_{\rm asym}|~$& $10$ &
    $10^{-1}$\\
  \hline
   & &\\
  $|C_{\rm {TE_{1,0}}}|$ & $1$ &
  $10^{-1}$ \\
  \hline
\end{tabular}
\caption{ Order of magnitude of the
absolute value of the C factor for all trajectories, for
both dipoles and electrons.} \label{Tabla}
\end{table}

In electron's
interference experiments, we conclude that the best experimental
setup would be either the asymmetric trajectory or the elliptic path.
In those configurations, for adequate parameters of the trajectories
it is in principle possible to obtain a complete destruction of the
interference pattern (setting the value of $\vert C\vert$ equal to
a zero of the Bessel function $J_0$).
In an interference experiment with dipoles, the best experimental
setup would be either the asymmetric trajectory or the waveguide.
In these cases, the effect is non negligible thanks to the fact that
the $C$ factor is independent of the velocity.
Moreover, as shown in the Appendix,
one is allowed to increase the intensity of the external field,
since the scattering cross section for dipoles is much lower than
for electrons, being still possible to neglect the direct interaction
with the electromagnetic field.


\section{Final Remarks}

We have estimated the loss of contrast produced when interfering
particles are shined by a classical electromagnetic field.
We considered both a monochromatic, linearly polarized electromagnetic field,
and electromagnetic fields inside a waveguide. We
considered different trajectories for both charged particles and dipoles.
Symmetric and asymmetric paths have been used in order to illustrate the
dependence of the fluctuating phase upon the velocity.
We have shown examples (asymmetric trajectories, waveguide) in which
the loss of contrast for dipoles is independent of the velocity in the low
velocity limit. This particular behaviour does not show up for
charged particles. However, we have found that
the effect for electrons is in general larger than the one computed
in Ref. \cite{ford} for a particular symmetric trajectory.

We also
estimated the backreaction effect of the fields over the dipoles,
evaluating the scattering cross section. When the mean free path for
dipoles is much larger than
the characteristic dimensions of their trajectories, the
direct interaction between dipoles and photons can be
neglected. This condition puts an upper bound over the external
field intensity.
Given an external field, the decoherence for dipoles is smaller
than the one expected for electrons. However, as the scattering
cross section is also smaller, it is in principle possible to increase the
intensity of the  external fields in order to partially compensate the
difference, still within the upper bound mentioned above.

Finally, we estimated the magnitude of the effect for values
of the different parameters achievable in the laboratory.
In the case of electrons, for an adequate setup the interference
fringes can dissappear totally. For dipoles, the
independence with the velocity makes the effect much more
important than naively expected, and could be observed
for sufficiently strong external fields.

\section{Appendix}

The loss of contrast computed in the previous sections depends on
the intensity of the classical electromagnetic field. In the usual
AB and AC effects, the force on the charges and dipoles vanishes.
However, in the time dependent case considered in this paper, the
dipoles have a direct interaction with the electromagnetic field,
which we neglected. Therefore the intensity of the electromagnetic
field is limited by the scattering cross section: the mean free
path for dipoles should be much larger than the characteristic size of
its trajectory.

In this section we will evaluate the scattering cross section for
both dipoles in interaction with a plane wave, and dipoles
travelling inside the waveguide, and compare these with the
results for charged particles.

\begin{enumerate}

\item Plane Wave.
The force experienced by a non-relativistic neutral atom with
arbitrary electric and magnetic dipole moments in the presence of
an electromagnetic field is \cite{schwinger} \beq
\mbf{F}=\mbf{\nabla} \bigg[ \mbf{d} \cdot \mbf{E}(\mathrm{R}) +
\mbf{m} \cdot \mbf{B}(\mathrm{R}) \bigg] + \partial_t \big(
\mbf{d} \times \mbf{B}(\mathrm{R}) \big), \eeq where $\mathrm{R}$
is the position of the center of mass of the atom.

For the plane wave of Section II, the force on the particle reads
\beq \mbf{F}=-k_y E_0 \cos(\omega t-k_y y) (d_y~\hat{z}
+m_x~\hat{y}). \eeq

If the atom's center of mass is oscillating around the origin of
coordinates ($y=0$), we can approximate the force it feels as the
force evaluated at $y=0$ for every time $t$. In the
non-relativistic limit, the acceleration of the particle is
$\ddot{\mbf{x}}=\mbf{F}/m_{\rm A}$. Writing the dipole moment as $\mbf{d}=
e \mbf{x}$, we can compute $\mbf{\ddot{d}}$ from the force, and
use Larmor's formula to know the angular distribution of radiated
power $\frac{\mbf{d}P} {d {\mbf{\Omega}}}=(1/4
\pi)|\mbf{\ddot{d}}|^2 \sin^2 (\Theta)$,
where $\Theta$ is the angle
between $\mbf{\ddot{d}}$ and $\mbf{n}$, a unit vector. The
scattering cross section, averaged in time, is therefore \beq
\sigma_d= \frac{8 \pi}{3}\frac{e^2}{m_{\rm A}^2} k_y^2 (d_y^2+m_x^2). \eeq

We can compare this result with Thomson's scattering cross section
$\sigma_e= \frac{8 \pi}{3} (\frac{e^2}{m_e})^2$ for the case of a
non-relativistic electron interacting with a plane wave. If we
consider that $|d_y|>|m_x|$, then: \beq \sigma_d \approx \sigma_e
(\frac{L}{\lambda})^2 (\frac{m_e}{m_{\rm A}})^2, \eeq where $L$ is the dipole's
characteristic length and $\lambda$ is the wavelength of the
field. Thus, we can see that the mean free path for dipoles
$l_{\rm mfp}^d \sim 1/\sigma_d$ is bigger than the corresponding
$l_{\rm mfp}^e$ for electrons in a factor $(\lambda/L)^2( m_{\rm A}/ m_e)^2$.

Based on this simple observation, we conclude that the suppression
of fringe visibility in the experiment done with coherent
electrons or with coherent dipoles could have a similar order of
magnitude. Indeed, although for a given external field the
suppression is bigger for electrons than for dipoles, in the
latter case it is possible to increase the intensity of the
external field to partially compensate the difference, still within the
limit of negligible direct interaction.

\item Waveguide.
For the case of a dipole interacting with the fields of the TE mode inside a
waveguide the force is written as:
\beqa \mbf{F}_{d}=\frac{k k_z}{\gamma^2}  B_0 \bigg[d_x k_x  \sin (k_x x) \sin (k_z z)
\sin (k_y y - \omega t) - d_y  k_y \cos(k_x x) \sin(k_z z) \cos (k_y y - \omega t)
\nonumber \\
- d_z k_z \cos(k_z z) \cos (k_x x) \sin (k_y y -  \omega t ) \bigg] \hat{x} +
\frac{k k_x}{\gamma^2} B_0 \bigg[ d_x k_x  \cos(k_x x) \cos (k_z z)
\sin (k_y y - \omega t) \nonumber \\
+ d_y k_y \sin(k_x x) \cos(k_z z) \cos(k_y y - \omega t)- d_z  k_z \sin(k_x x)
\sin(k_z z) \sin (k_y y - \omega t)\bigg] \hat{z}
 \eeqa
Following the reasoning above for the case of the plane wave, we
obtain that the scattering cross section of the TE mode is given
by  \beq \sigma_{\rm TE}^d \approx \frac{8 \pi}{3} \frac{e ^2
d_y^2}{m_{\rm A}^2} k k_y = \frac{8 \pi}{3} \frac{e^4}{m_{\rm A}^2} L^2 k k_y =
\sigma_{\rm TE}^e k_y ^2 L^2 (\frac{m_e}{m_{\rm A}})^2
 \eeq
where $\sigma_{\rm TE}^e=\frac{8 \pi}{3} \frac{e^4}{m_e^2}
\frac{k}{k_y}$ is the total cross section for an electron inside
the waveguide in the TE mode (for simplicity we assumed that
$k_x \sim k_z$, and $a \sim b$). Therefore, we obtain
$\sigma_{\rm TE}^d \ll \sigma_{\rm TE}^e$, and  $l_{\rm mfp}^{d,\rm TE} \gg
l_{\rm mfp}^{e,\rm TE}$

For the TM modes, an analogous calculation shows that
$\sigma_{\rm TM}^d\approx \sigma_{\rm TE}^d$.

\end{enumerate}

\section{Acknowledgments}
We thank Juan Pablo Paz for useful discussions.
This work was supported by UBA, CONICET, Fundaci\'on Antorchas, and
ANPCyT, Argentina.

\end{document}